\documentclass[a4paper]{spie}

\usepackage{amsmath,amsfonts,amssymb}
\usepackage{graphicx} 
\usepackage{epstopdf}
\usepackage{siunitx}
\usepackage{subcaption}
\usepackage[colorlinks=true, allcolors=blue]{hyperref}

\title{Optical atomic magnetometry for magnetic induction tomography of the heart}

\author{Cameron Deans}
\author{Luca Marmugi}
\author{Sarah Hussain}
\author{Ferruccio Renzoni}
\affil{Department of Physics and Astronomy, University College London, Gower Street London WC1E 6BT, United Kingdom}

\authorinfo{Send correspondence to Luca Marmugi, e-mail: l.marmugi@ucl.ac.uk. \\ Pre-print version. Article reference: Proc. SPIE \textbf{9900}, Quantum Optics, 99000F (April 29, 2016); \href{http://dx.doi.org/10.1117/12.2227538}{DOI: 10.1117/12.2227538}.}

\begin{document} 
\maketitle

\begin{abstract}
We report on the use of radio-frequency optical atomic magnetometers for magnetic induction tomography measurements. We demonstrate the imaging of dummy targets of varying conductivities placed in the proximity of the sensor, in an unshielded environment at room-temperature and without background subtraction. The images produced by the system accurately reproduce the characteristics of the actual objects. Furthermore, we perform finite element simulations in order to assess the potential for measuring low-conductivity biological tissues with our system. Our results demonstrate the feasibility of an instrument based on optical atomic magnetometers for magnetic induction tomography imaging of biological samples, in particular for mapping anomalous conductivity in the heart.
\end{abstract}


\section{Introduction}
\label{sec:intro}

There is currently no technique capable of mapping the conductivity of biological tissues. Widely used approaches, for example magnetocardiography and electrocardiography, do not provide any direct information regarding conductivity. Nevertheless, the imaging of conduction throughout the body, in particular across the heart, is important for the understanding of many diseases, such as cardiac arrhythmias.

Here, we report on the ongoing research towards the development of a diagnostic tool for investigations into the conductivity of living tissue, based on Magnetic Induction Tomography  (MIT) performed with Optical Atomic Magnetometers (OAMs)\cite{marmugi2016}. We couple the advantages of MIT with the performance of  OAMs\cite{apl_paper, wickenbrock2014}. The overall performance of an MIT system is dependent on the characteristics of the magnetic field sensor used: conventional MIT systems have limited sensitivity at low frequency, limited bandwidths, and have spatial resolution limited by their size. OAMs overcome these limitations owing to their potential for extreme sensitivity and miniaturization. Furthermore, our system operates at room temperature in an unshielded environment. Based on these facts, we envisage an instrument capable of ultra-sensitive high-resolution imaging, using arrays of atomic magnetometers operating in a MIT modality.

\subsection{Magnetic Induction Tomography}
\label{sec:MIT}

MIT allows the non-invasive investigation of the passive electromagnetic properties of materials\cite{mit}.  A primary AC magnetic field, $\mathbf{B_1}$, is applied to induce eddy currents in the object of interest. The eddy currents generate an additional magnetic field component, $\mathbf{B_2}$. The density of these currents, and hence the magnitude of $\mathbf{B_2}$, are determined by the electromagnetic properties of the target object. In particular, the response is determined by the electrical conductivity ($\sigma$), the relative permittivity ($\epsilon_{r}$), and the relative permeability ($\mu_{r}$)\cite{miteqn}.

\subsection{Optical Atomic Magnetometry}
\label{sec:OAM}

OAMs aim to determine the properties of an unknown magnetic field by measuring its influence on an atomic vapor. This is done by exploiting light-matter interactions with a resonant probing laser beam. Generally, there are three fundamental stages in optical atomic magnetometry.

The first stage is the preparation of macroscopic spin-polarization in the sample. This is performed by ground-state optical pumping. A static magnetic field is used to impose a quantization axis in the vapor. A circularly polarized pump beam propagates through the sample along this axis. This beam is tuned to drive an optical transition, for an alkali vapor this transition is the $D_{1}$ or $D_{2}$ line. The pump beam transfers angular momentum between photons and the atoms' Zeeman sub-levels. This results in the accumulation of the atomic population in the sub-level with maximum projection along the quantization axis, creating a quantum spin-state that is sensitive to external magnetic fields.

In the second stage, the atomic polarization evolves in the surrounding unknown magnetic field that is to be measured. In the additional magnetic field, a torque acts on the atomic sample causing the angular momentum to precess around the resultant total magnetic field. This occurs at the Larmor frequency, $\omega_{L} = \gamma \lvert \mathbf{B_{TOT}} \rvert$, where $\gamma$ is the gyromagnetic ratio and $\mathbf{B_{TOT}}$ is the total magnetic field experienced by the vapor. This reduces the measurement of the magnetic field to a measurement of the Larmor precession. 

The third stage is to perform this measurement with a linearly polarized probe beam. The spin-polarization of the vapor interacts with the probe beam via the Faraday effect. The plane of polarization of the probe beam is periodically modulated at $\omega_{L}$. Hence, by measuring the polarization rotation, the magnetic field properties can be inferred.

\subsubsection{Radio-frequency OAMs}
\label{sec:RFOAM}

RF optical atomic magnetometers (RF-OAMs) act on two different transitions present in the alkali atomic sample used \cite{rfoam1, rfoam2, rfoam3}. In addition to the hyperfine transitions, used for the optical pumping and detection of spin precession, an RF magnetic field is applied to coherently drive the populations of the magnetic sub-levels within the ground-state structure. These transitions create an oscillating transverse spin polarization component that is transferred to rotations in the polarization of the probe beam. Hence, the magnitude and phase of the oscillations in the polarization can be used to measure the corresponding properties of the RF magnetic field.

We perform MIT measurements using an RF-OAM as the magnetic field sensor \cite{apl_paper}. This approach has a number of advantages over conventional MIT systems, which use pick-up coils to to detect the magnetic field perturbations\cite{marmugi2015}.

\subsection{Medical applications of MIT with OAMs}
\label{sec:medical}

The fact that the bandwidth of OAMs is not limited by intrinsic factors is crucial for applications in biomedical imaging \cite{medapp1, medapp2, medapp3}. This allows for the detection across the large frequency range required for eddy current excitations in low-conductivity targets, such as biological tissues. Another obvious advantage for medical applications is the completely non-invasive approach. 

Our OAM-based MIT system has the potential to fulfil these requirements: in Section~\ref{sec:comsol}, we explore the the optimum working point of our device for biomedical applications via finite element simulations. The instrument finds immediate applications in the investigation and diagnosis of conditions for which changes in conductivity are present\cite{marmugi2016}. These include cardiac arrhythmias, such as atrial fibrillation (AF), which manifest as irregular beatings of the heart.

The fundamental causes of AF, which affects more than 10\% of the population over 70, are little understood. Indications suggest that AF is caused by permanent changes in the local conductivity of the heart, producing deterministic sources known as rotors \cite{rotors1, rotors2}. Though the subject remains hotly debated \cite{rotors3}. A conductivity map of the heart would, therefore, shed light both on the fundamental causes of AF and help provide effective diagnosis and improve clinical treatments. Among the other potential medical applications is imaging the electric properties of malignant tissues. Conductivity of specific cancerous tumors has been shown to differ greatly from that of the surround tissues. Hence, a conductivity map of the region could prove to be an effective tool in the early detection and diagnosis of tumors \cite{marmugi2016}.

\section{Experimental Set-up}
\label{sec:setup}

The detailed description of our RF-OAM set-up has recently been reported \cite{apl_paper}. Here, we summarize the essential points. The sensor is a \SI{25}{\mm} cubic cell containing an isotopic mixture of Rb vapor and N$_{2}$ as a buffer gas. A uniform static magnetic field is applied in the $\hat{z}$ direction. Ground state spin-polarization along $\hat{z}$ is performed by optical pumping via a $\sigma^{+}$ polarized pump beam tuned to the $D_{2}$ line $F=2 \rightarrow F'=3$ transition of $^{87}$Rb, or to the $F=3 \rightarrow F'=4$ transition of $^{85}$Rb as described in Section~\ref{subsec:comparison}. An AC magnetic field, applied along $\hat{y}$, drives both the RF-OAM and acts as the primary field ($\mathbf{B_1}$) for the MIT. With a target object present, the secondary magnetic field ($\mathbf{B_{2}}$) is induced due to the flow of eddy currents. The total magnetic field, $\mathbf{B_{TOT}}=\mathbf{B_{1}}+\mathbf{B_{2}}$, that is felt by the sensor causes coherent population transfer in the atomic ground-state sub-levels. This results in an oscillating transverse atomic polarization. A $\pi$-polarized probe beam, detuned by +\SI{425}{\mega\hertz}, propagates through the sensor in the $\hat{x}$ direction. The plane of polarization of this beam is rotated by the Faraday effect and is measured via a balanced polarimeter. The output of the polarimeter is sent to a lock-in amplifier, referenced to the the driving field $\mathbf{B_{1}}$. This gives both the radius (R) and phase ($\phi$) of the Faraday rotation signal produced by the total magnetic field. This field varies with the contribution $\mathbf{B_2}$ as the object is moved around the sensor on a translational stage, allowing position-resolved R and $\phi$ maps of target objects.

\section{Results}
\label{sec:results}

\subsection{Towards Imaging of Low Conductivity Objects}
We recently demonstrated the MIT imaging of conductive objects using a RF-OAM \cite{apl_paper}.  However, the conductivity approaches \SI{1}{\siemens\per\metre} for applications in biomedical imaging.

Here, we explore how the imaging performance of our system is affected by changes in the conductivity of the target object. In order to isolate the conductivity's contribution, objects of the same size and geometry are analyzed in the same experimental conditions.

We image homogeneous samples of copper ($\sigma_{\text{Cu}}=$~\SI{5.98e7}{\siemens\per\metre}) and manganese ($\sigma_{\text{Mn}}=$~\SI{5.40e5}{\siemens\per\metre}), both squares of dimensions \SI{25}{\milli\metre}$\times$\SI{25}{\milli\metre}$\times$\SI{1}{\milli\metre}. In the case of conductive media, the field's penetration in the bulk is exponentially attenuated with a space constant $\delta(\omega)$ given by\cite{skindeptheqn}:

\begin{equation}
\delta(\omega)=\sqrt{\dfrac{2}{\omega \mu \sigma}}~,\label{eqn:skindepthhf}
\end{equation}

\noindent where $\mu=\mu_{0} \mu_{r}$ is the magnetic permeability.

Given Eq.~\ref{eqn:skindepthhf}, we tune the RF-OAM accordingly in order to obtain $\delta=$~\SI{1.5}{\milli\metre} for both samples. Hence, the ratio between skin depth and sample's thickness ($\delta/h=$~1.5) is held constant. For copper, the primary field's frequency is $\omega^{Cu}/2\pi=$\SI{1.88}{\kilo\hertz}; for manganese, $\omega^{Mn}/2\pi=$~\SI{208.48}{\kilo\hertz}.

\begin{figure}[htbp]
\centering
\includegraphics[height=10cm]{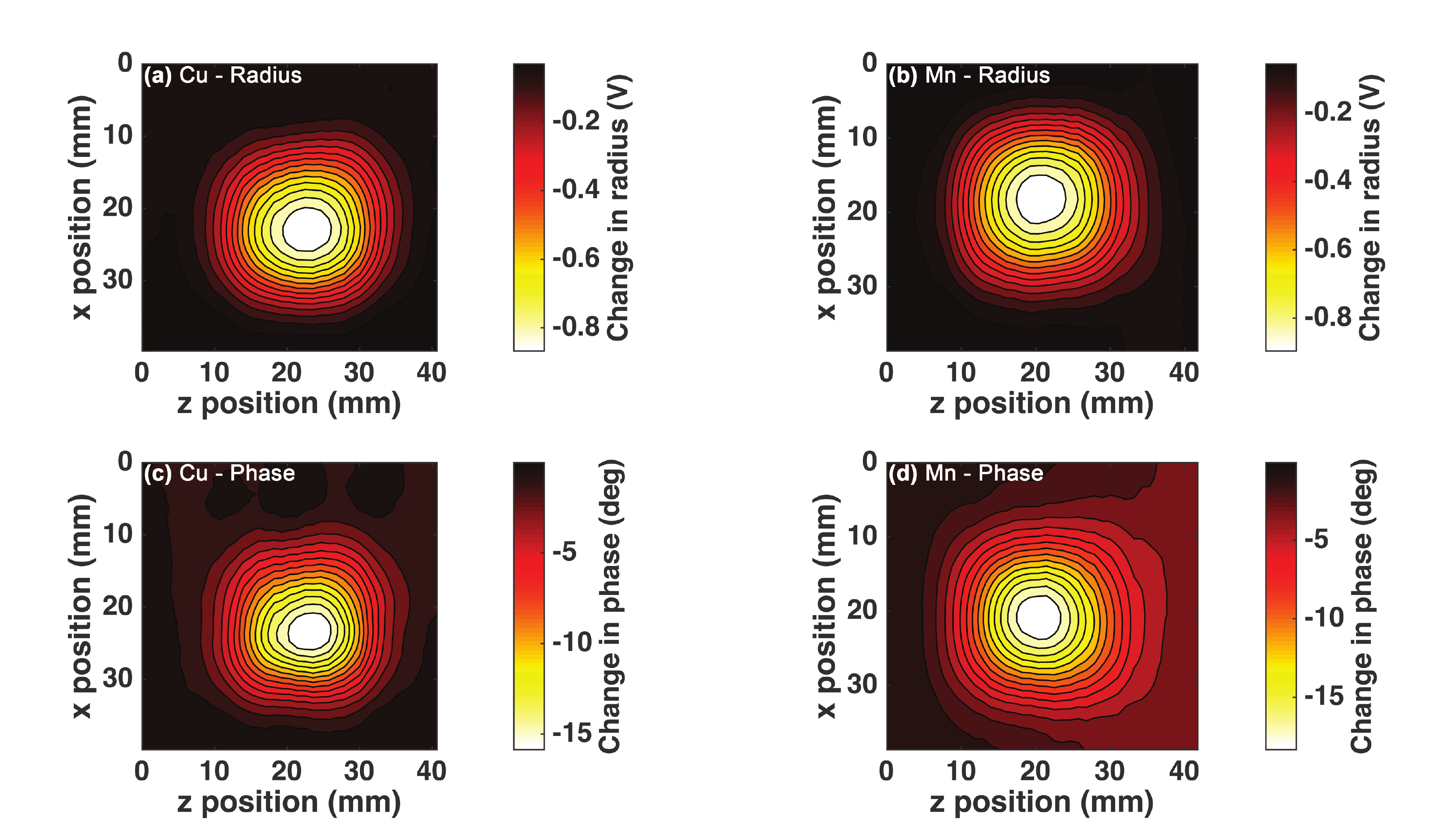}
\caption{\textbf{Magnetic induction imaging with an RF-OAM of Cu and Mn samples.} \textbf{(a), (c)} Contour plots of the radius  (a) and of the phase (c) maps of a Cu square (\SI{25}{\milli\metre}$\times$\SI{25}{\milli\metre}$\times$\SI{1}{\milli\metre}) at  \SI{1.88}{\kilo\hertz}.  \textbf{(b), (d)} Contour plots of the radius (b) and the phase (d) maps of a Mn square (\SI{25}{\milli\metre}$\times$\SI{25}{\milli\metre}$\times$\SI{1}{\milli\metre}) at  \SI{208}{\kilo\hertz}. In all cases, the ratio between the skin depth and the samples' thickness is 1.5.}
\label{fig:cumn}
\end{figure}

These conditions are chosen to allow immediate evaluation of different images. We present in Figure \ref{fig:cumn} the direct comparison of the Cu and the Mn samples. Our system is capable of a large dynamic range, allowing imaging of samples with conductivities differing by two orders of magnitude. Response and accuracy of objects' reproduction are satisfactory in every case, with the shape and the size of the object not affected by the conductivity of the sample. 

The similarity of the measurements is not surprising: given the 1.5 ratio between the skin depth and the sample's thickness, similar conditions for the primary field's penetration are created. These results demonstrate the flexibility of our system, mainly thanks to the superior bandwidth and sensitivity of the OAM. In particular, imaging without significant performance degradation is performed in a band of more than \SI{200}{\kilo\hertz} and with a conductivity spanning from \SI{1e5}{\siemens\per\metre} to \SI{1e7}{\siemens\per\metre}.

\subsection{Increasing the contrast: $^{87}$Rb and $^{85}$Rb OAMs}\label{subsec:comparison}
The quest for the optimal working conditions for imaging of biological tissues leads to improvement in the sensitivity and, consequently, in the overall contrast of the conductivity maps produced via MIT.

The sensitivity of an OAM depends on the number of spins (i.e. polarized atoms) involved in the quantum measurement. In particular, the effective minimum detectable magnetic field for an optical magnetometer operating with alkali atoms is proportional to $1/\sqrt{N}$, where $N$ is the number of polarized atoms \cite{oambook}. Therefore, we also tested the performance of our system with $^{85}$Rubidium. Given the different natural abundance of the two isotopes ($^{87}$Rb: 28\%; $^{85}$Rb: 72\%), an increase in the MIT signal can be expected. It is noteworthy that the 85 isotope could be more prone to atomic losses due to hyperfine optical pumping. Furthermore, with increasing temperature, its larger atomic density could in principle produce detrimental effects because of spin-depolarizing collisions and optical thickness of the vapor. Nevertheless, none of these effects was observed in our experimental conditions.

\begin{figure}[htbp]
\centering
\includegraphics[height=7cm]{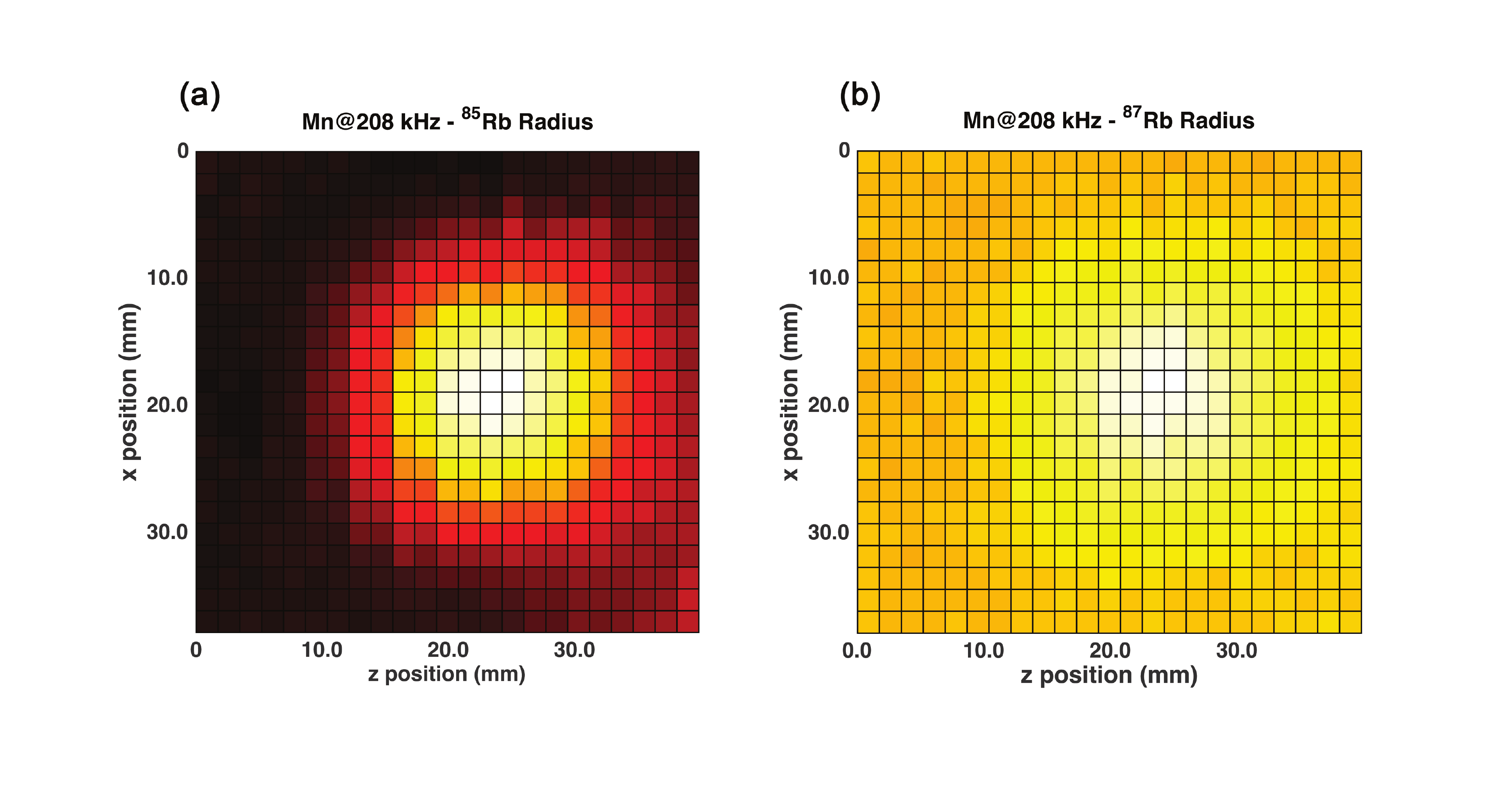}
\caption{\textbf{MIT imaging of a Mn sample with an  $^{85}$Rb (a) and  $^{87}$Rb (b) OAM.} Radius maps, normalized according to Eq.~\ref{eqn:normalisation}, obtained in the same operational conditions at \SI{208}{\kilo\hertz}.}
\label{fig:comparison}
\end{figure}

Direct comparison in the case of the Mn square imaged at  \SI{208}{\kilo\hertz} is presented in Figure \ref{fig:comparison}. It is noteworthy that the only changes were the tuning of laser and the corresponding adaptation of the static magnetic field along the pump beam to the different properties of the $^{87}$Rb and $^{85}$Rb magnetic states.

In order to highlight the difference in performance, the absolute value of the relative variation with respect to the $^{85}$Rb case is computed at each pixel $(x,z)$:

\begin{equation}
R_{xz}=\left|\dfrac{r_{xz}^{(85, 87)}-min(r_{xz}^{(85, 87)})}{max(r_{xz}^{(85)})}\right| \label{eqn:normalisation}
\end{equation}

\noindent and plotted in the matrices of Figure \ref{fig:comparison}. No nearest-neighbor averaging is applied here, so as to allow a more direct comparison.

Comparison between part (a) and (b) of Figure \ref{fig:comparison} reveals a clear improvement in the images' contrast when using the more abundant $^{85}$Rb, in the case of a relatively low conductive metal. This demonstrates a viable path for improving the overall performance of the OAM-based system in view of the proposed applications to biological tissues and, in particular, the diagnosis of heart's conductivity anomalies. 

\begin{figure}[htbp]
\centering
\includegraphics[height=5.5cm]{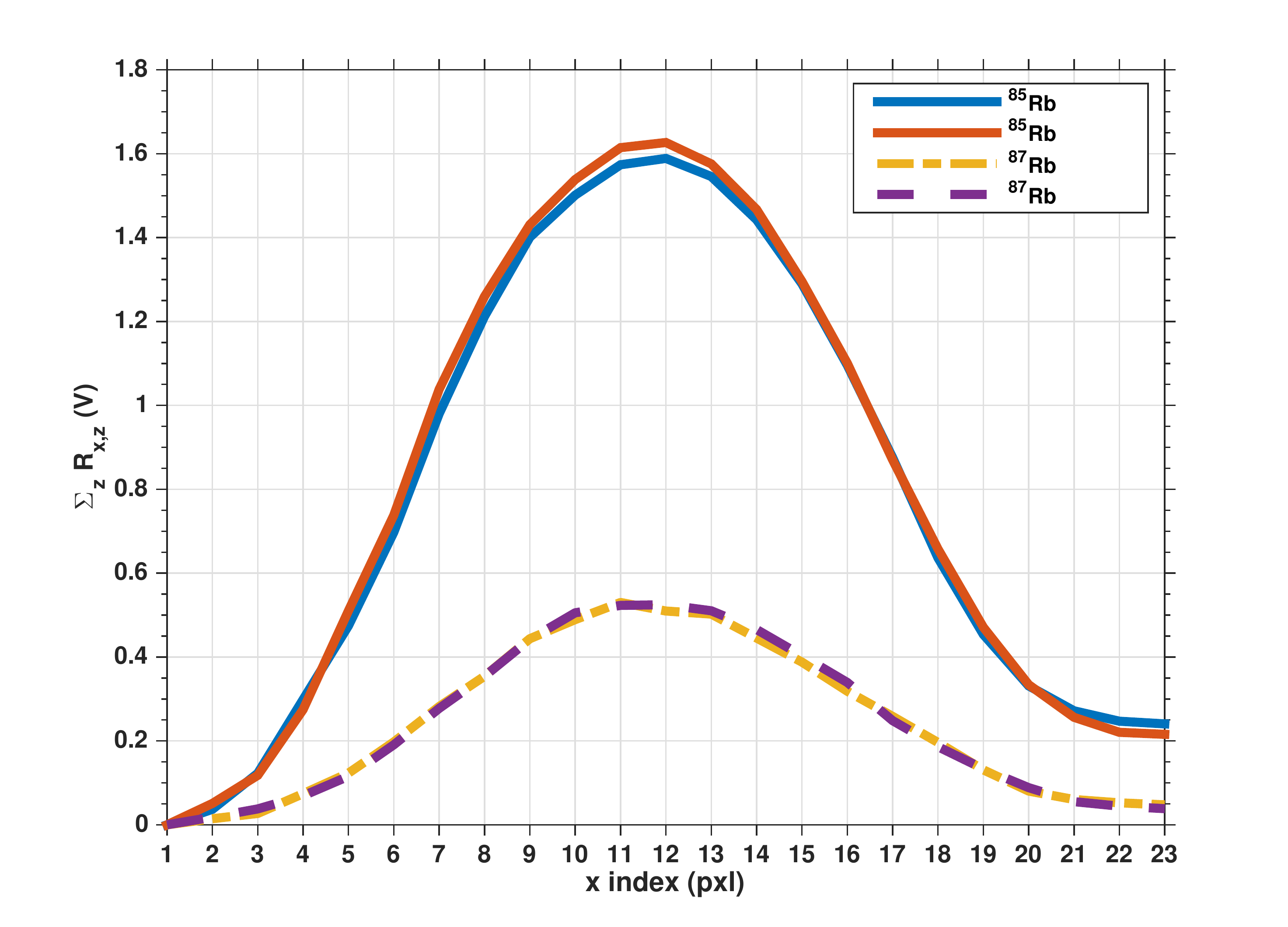}
\caption{\textbf{Cumulative total radius change measured with a Mn sample with an $^{85}$Rb and  $^{87}$Rb RF-OAM}. Sum over rows of the absolute radius variation measured at \SI{208}{\kilo\hertz}. The blue continuous line ($^{85}$Rb) and the dashed/dotted yellow line ($^{87}$Rb) correspond to panel (a) and (b) of Figure \ref{fig:comparison}, respectively.}
\label{fig:xsection}
\end{figure}

In order to provide a more quantitative evaluation of the performance improvement, values obtained during the comparative test are summed per rows and plotted as a function of pixels' index (Figure \ref{fig:xsection}). The binning clearly shows the different variations measured with the two isotopes and the consistency of the outcome: all the images of the Mn sample realised with the 85 isotope exhibit a factor $\sim$3 improvement with respect to those obtained with the 87 isotope.

\section{MIT of anomalous conduction in the heart: simulations}
\label{sec:comsol}

In order to assess the feasibility of medical imaging with our device, we perform finite element method simulations using \textit{COMSOL multiphysics}. In particular, we consider the application of producing a conductivity map of the heart \textit{in vivo}. 

\begin{figure}[htb]
\begin{center}
\begin{tabular}{c}
\includegraphics[height=4cm]{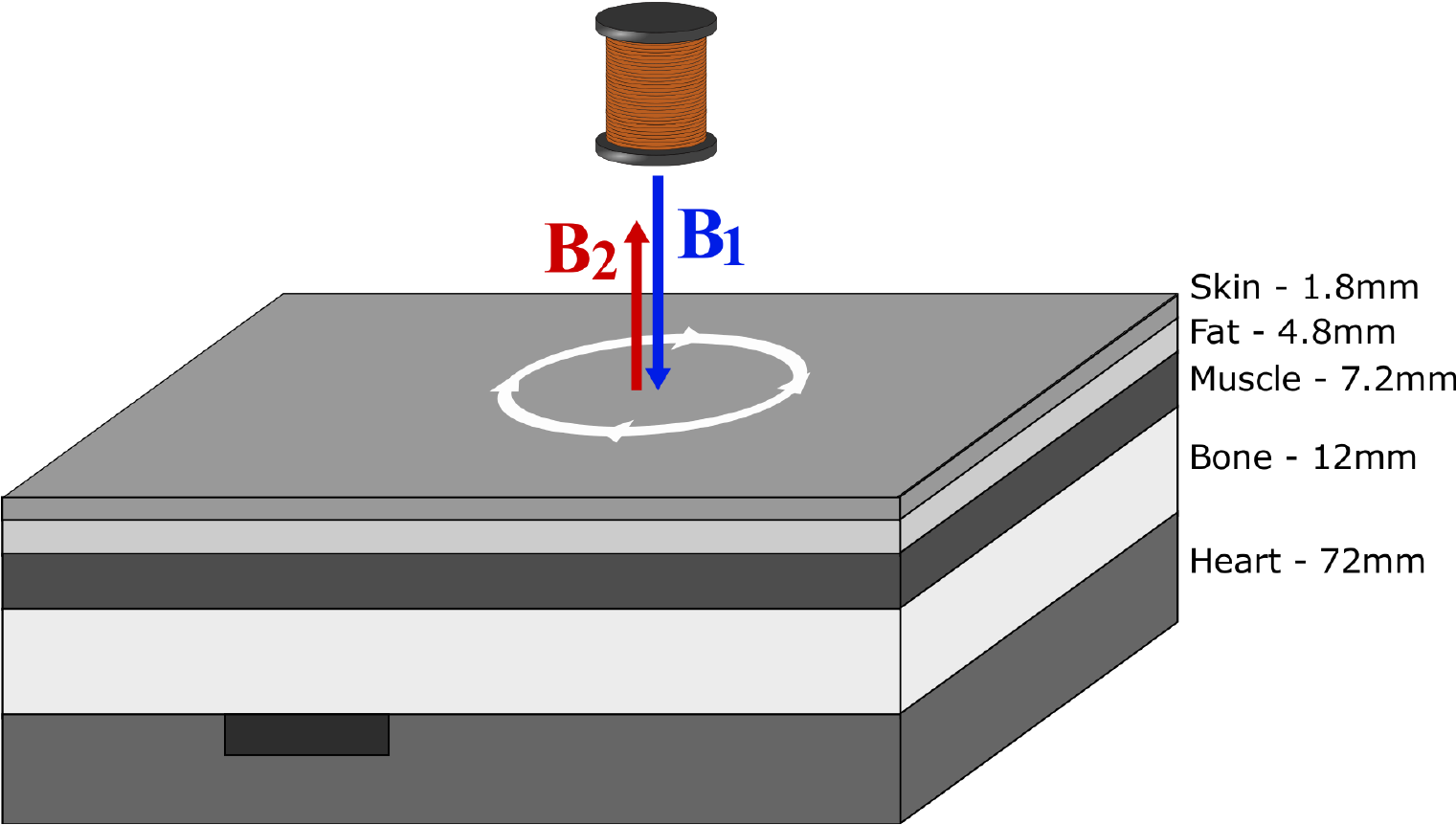}
\end{tabular}
\end{center}
\caption{Schematic of the model used in finite element simulations, not to scale. The coil is an exact model of the one in our experimental set-up. Electromagnetic properties and thickness's of each layer are set to their expected values. A conductivity anomaly mimicking the source of cardiac fibrillation (\textit{arrhythmogenic locus}) at the surface of the heart layer is represented in black.}
\label{fig:layer_model}
\end{figure}

A schematic of the model used is shown in Figure~\ref{fig:layer_model}. The excitation coil for the primary field is an exact model of the one used in our experimental set-up (\SI{7.8}{\milli\metre} diameter, \SI{580}{\micro\henry}). The coil is positioned above a layered model representing the different tissues that surround the heart. The thickness of each layer was set in line with known values and the electronic properties were calculated following methods described in the literature\cite{tissueproperties1, tissueproperties2, marmugi2016}. We insert an additional region whose conductivity is varied independently, representing a conductivity anomaly and hence a potential \textit{arrhythmogenic locus}, at the surface of the heart layer. In each simulation the excitation frequency of the coil is set and we record the magnetic field produced. This field includes the contribution from the eddy currents from the model. We record the magnetic field across a line at a distance of \SI{6}{\milli\metre} above the surface on the skin and a displacement of \SI{10}{\milli\metre} from the coil. This position is chosen as it represents a possible positioning of the RF-OAM sensor. 

The secondary field response is dependent on both the excitation frequency and the conductivity:
 
\begin{equation}
\frac{\mathbf{B_{2}}}{\mathbf{B_{1}}} = A \omega \mu_{0} \left[ \omega	\epsilon_{0} (\epsilon_{r} - 1 ) - i\sigma \right] + B(\mu_{r}-1)~, 
\label{miteqn}
\end{equation}

\noindent where $\epsilon_{0}$ and $\mu_{0}$ are the permittivity and permeability of free space and A and B are geometric constants\cite{miteqn}.

For biological samples, where $\mu_{r}=1$, the real part contribution scales with $\omega^{2}$ whereas the imaginary contribution scales with $\omega\sigma$. Therefore, for low conductivity samples, a higher frequency allows both a increased secondary field and an increased response to conductivity changes. In addition, the penetration depth of a high frequency excitation field through a low-conductivity material is given by \cite{skindeptheqn},

\begin{equation}
\delta(\omega) = \frac{1}{\omega} \left[ \frac{\mu \epsilon_{r} \epsilon_{0}}{2} \left( \sqrt{1+\left(\frac{\sigma}{\omega \epsilon_{r} \epsilon_{0}} \right)^{2}} - 1 \right) \right]^{-\frac{1}{2}}.
\label{skindepth}
\end{equation}

\noindent Hence, for biomedical applications of MIT there is a trade-off between an increased secondary field response and a decrease in primary field penetration, at higher frequencies.

\begin{figure} [htb]
\includegraphics[width=0.91\linewidth]{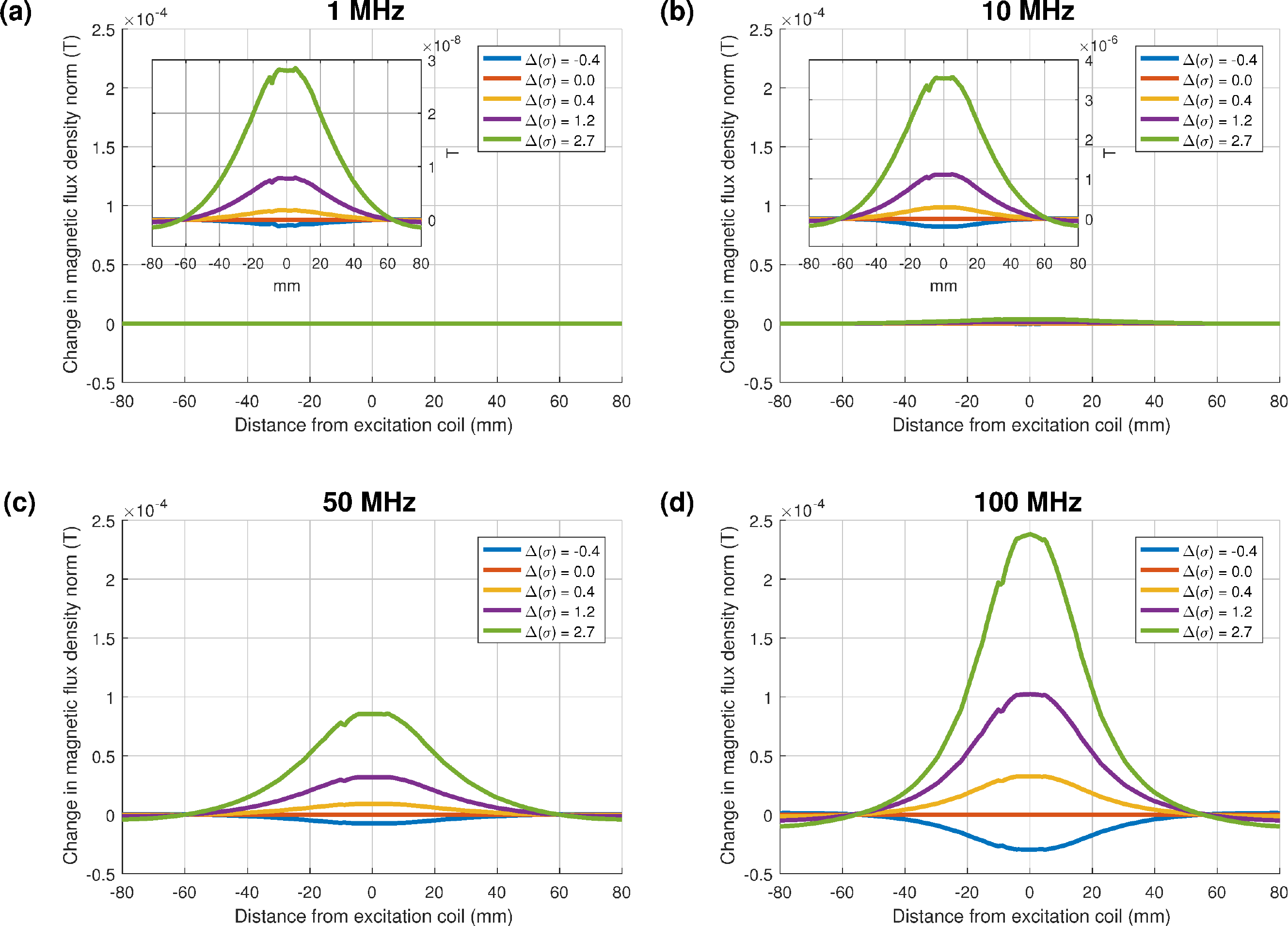}
\vspace{0.3cm}
\caption{Magnetic field change due to changing conductivity in the heart layer. Primary field excitation; \textbf{(a)}~$=$~\SI{1}{\mega\hertz}, \textbf{(b)}~$=$~\SI{10}{\mega\hertz}, \textbf{(c)}~$=$~\SI{50}{\mega\hertz}, \textbf{(d)}~$=$~\SI{100}{\mega\hertz}. At each frequency the conductivity of the heart layer is changed by a value $\Delta(\sigma)$ from the expected value, with all other properties are held constant. Main plots show the resulting change in the magnetic field along a sensing line \SI{6}{\milli\meter} above the skin and displaced by \SI{10}{\milli\metre} from the coil, against the same scale for comparison. For clarity, the insets display an expanded view of the field change.}
\label{fig:heart}
\end{figure}

We explore this trade-off by simulating the dependence of the magnetic field change on both the excitation frequency and the conductivity of the heart layer (Figure~\ref{fig:heart}). At each frequency, the conductivity of the heart layer is slightly modified and the change in the magnetic field at the sensing level is computed. As expected, the magnetic field increases with increasing conductivity and decreases with decreasing conductivity. Each of the plots is plotted against the same scale to allow comparison of the effect changing the frequency has on the response. The secondary field contribution is shown to increase with frequency over the range \SI{1}{\mega\hertz} - \SI{100}{\mega\hertz}. Further investigations revealed that the change in the field decay in our model over this frequency range was not as significant as predicted by Eq.~\ref{skindepth}. 

These simulations show that a clear magnetic field change above the skin is produced by a small conductivity change ($\approx$\SI{1}{\siemens\per\meter}) in the heart in the \SI{}{\mega\hertz} regime. More importantly, at frequencies approaching \SI{100}{\mega\hertz}, the field change, of the order of \SI{1e-4}{\tesla}, is detectable using an RF-OAM in an unshielded environment at room temperature \cite{apl_paper}.

\begin{figure}[htb]
\begin{subfigure}{.5\textwidth}
  \hspace{1.6cm}
  \includegraphics[height = 5cm]{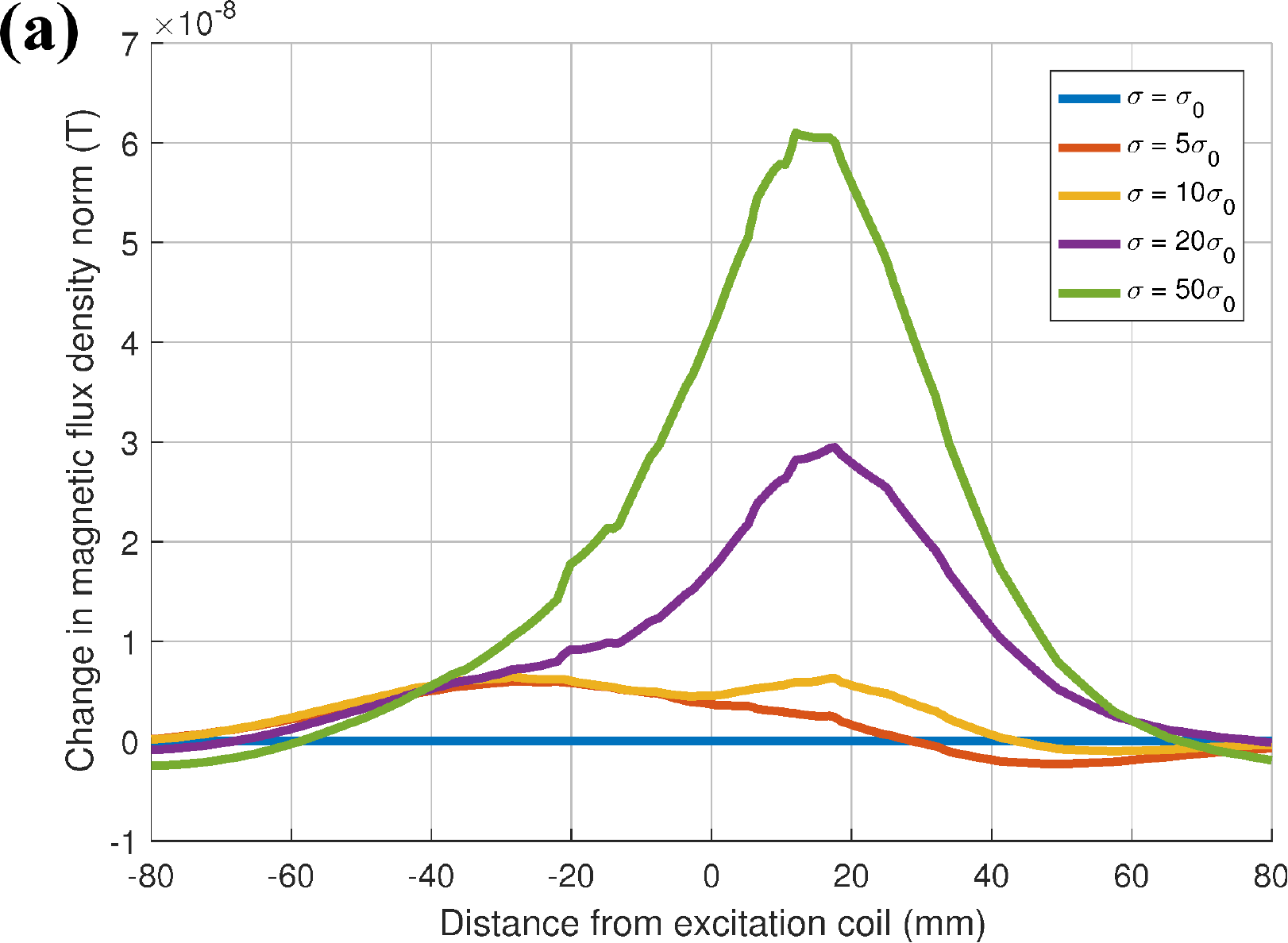}
\end{subfigure}
\begin{subfigure}{.5\textwidth}
  \hspace{0.7cm}
  \includegraphics[height = 5.3cm]{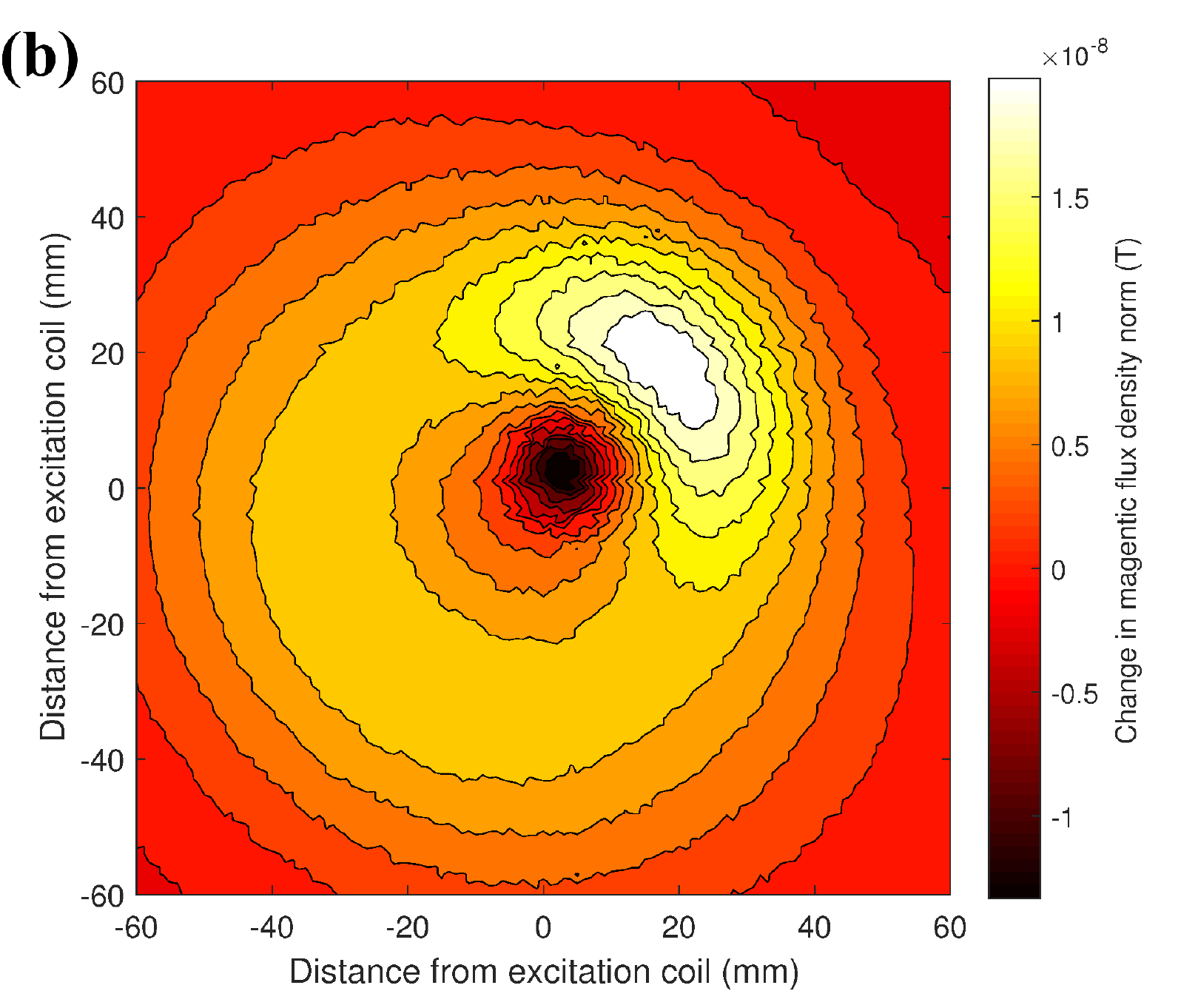}
\end{subfigure}
\begin{subfigure}{\textwidth}
  \centering
  \vspace{0.2cm}
  \includegraphics[width=.8\linewidth]{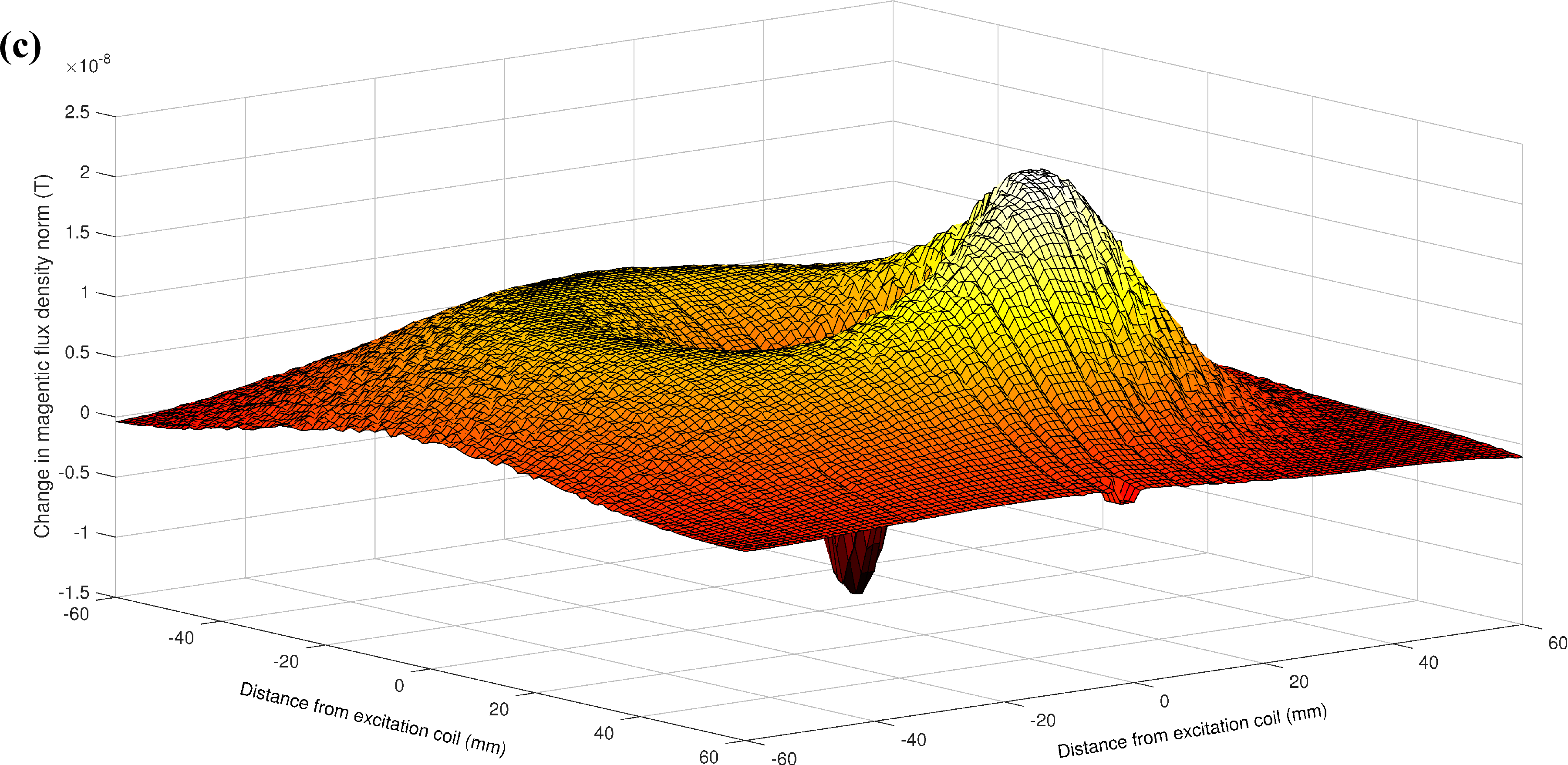}
\end{subfigure}
  \vspace{0.5cm}
\caption{Magnetic field change \SI{6}{\milli\meter} above the skin due to a conductivity anomaly at the heart's surface. \SI{10}{\milli\metre}$\times$\SI{10}{\milli\metre}$\times$\SI{5}{\milli\metre} anomaly centred at (\SI{10}{\milli\metre}, \SI{10}{\milli\metre}), primary field excitation \SI{100}{\mega\hertz}. \textbf{(a)} Magnetic field change above anomaly as the conductivity of the anomaly is varied ($\sigma_{0}=$~\SI{0.875}{\siemens\per\metre}). \textbf{(b)} and \textbf{(c)}, 2D contour plot and  3D rendering showing the magnetic field change in a slice for an anomaly conductivity of $\sigma = 20 \sigma_{0}$.}
\label{fig:anomaly}
\end{figure}

Furthermore, we simulate the application of an OAM based MIT system to the detection of conductivity anomalies in the heart (Section~\ref{sec:medical}) by varying the conductivity of a small region on the heart's surface (Figure~\ref{fig:anomaly}). This region is of size \SI{10}{\milli\metre}$\times$\SI{10}{\milli\metre}$\times$\SI{5}{\milli\metre} and is centered at (\SI{10}{\milli\metre}, \SI{10}{\milli\metre}) with respect to the excitation coil. A primary field frequency of \SI{100}{\mega\hertz} was chosen for these simulations, in view of the results above. 

In Figure~\ref{fig:anomaly}(a), we compute the magnetic field change along a sensing line above the anomaly and \SI{6}{\milli\metre} above the skin's surface. The conductivity of the anomaly is increased in multiples of $\sigma_{0}=$~\SI{0.875}{\siemens\per\metre} (the conductivity of the surrounding heart tissue). A peak in the magnetic field change is observed in the region above the anomaly in the regime $\sigma \geq 10\sigma_{0}$. Figures \ref{fig:anomaly}(b)/(c) display the magnetic field change in a slice above the skin's surface for $\sigma = 20\sigma_{0}$. A clear increase is found directly above the anomalous region. In particular, a maximum value of \SI{2.2e-8}{\tesla} is recorded at position (17,19). This local feature is the result of the additional excitation of local eddy currents due to the increased conductivity. In addition to this, a symmetric ring around the coil is observed. We attribute this general feature to the flow of eddy currents from the anomalous region around the coil position due to the symmetric shape of the magnetic field. The decrease in the centre of the ring is a result of the conservation of the total magnetic flux. This feature is common to all simulations.

\section{Conclusions}
We reported on the ongoing investigation towards the realisation of magnetic induction imaging of the heart and biological structures with optical atomic magnetometers. Magnetic induction tomography has already been identified as a promising technique for mapping the conductivity of living tissues. In particular, we demonstrated that our MIT system based on a RF-OAM can image low-conductive metallic objects. Thanks to the sensitivity, bandwidth and dynamic range of the magnetic sensor, comparable performance over \SI{200}{\kilo\hertz} and \SI{2}orders of magnitude of conductivity are demonstrated. Moreover, a clear improvement in the images' contrast can be obtained by using $^{85}$Rb. Finally, numerical simulations demonstrated the feasibility of \textit{in vivo} detection of anomalously conductive regions in the human heart, such as those responsible for heart fibrillation. These results represent a further step towards the realization of a novel class of diagnostic tools, with potential relevant impact in cardiology and in the wider biomedical field.

\acknowledgments  
This work was supported by a Marie Curie International Research Staff Exchange Scheme Fellowship ``COSMA'' (PIRSES-GA-2012-295264). C. D. acknowledges the support of the EPSRC Centre for Doctoral Training in Delivering Quantum Technologies. S. H. is supported by DSTL - Defence and Security PhD - Sensing and Navigation using Quantum 2.0 technology.


\end{document}